\documentclass[aps,preprint]{revtex4-1}
\usepackage{amsmath}
\usepackage{amssymb}
\usepackage{amsfonts}
\usepackage{color}
\usepackage[pdftex]{graphicx}

\newcommand\krho{p_{\rho}}

\newcommand\kz{p_z}
\newcommand\xhat{\mathbf{\hat{x}}}
\newcommand\yhat{\mathbf{\hat{y}}}
\newcommand\rhat{\mathbf{\hat{r}}}
\newcommand\lhat{\mathbf{\hat{l}}}
\newcommand\zhat{\mathbf{\hat{z}}}
\newcommand\kvec{\mathbf{k}}
\newcommand\rvec{\mathbf{r}}

\newcommand\Chkz{\mathbf{C}_{m\kz}}
\newcommand\Dhkz{\mathbf{D}_{m\kz}}

\newcommand\nhat{\mathbf{\hat{n}}}

\begin{document}

\title{Electromagnetic duality symmetry and helicity conservation for the macroscopic Maxwell's equations}

\author{Ivan Fernandez-Corbaton$^{1,2}$, Xavier Zambrana-Puyalto$^{1,2}$, Nora Tischler$^{1,2}$,  Alexander Minovich$^{3}$, Xavier Vidal$^1$, Mathieu L. Juan$^{1,2}$, and Gabriel Molina-Terriza$^{1,2} \email{gabriel.molina-terriza@mq.edu.au}$}

\affiliation{$^1$ QSciTech and Department of Physics \& Astronomy, Macquarie University, Australia}
\affiliation{$^2$ ARC Center of Excellence for Engineered Quantum Systems}
\affiliation{$^3$ Nonlinear Physics Centre, RSPE, Australian National University, Canberra, Australia}

\begin{abstract}
Modern physics is largely devoted to study conservation laws, such as charge, energy, linear momentum or angular momentum, because they give us information about the symmetries of our universe. Here, we propose to add the relationship between electromagnetic duality and helicity to the toolkit. Generalized electromagnetic duality symmetry, broken in the microscopic Maxwell's equations by the empirical absence of magnetic charges, can be restored for the macroscopic Maxwell's equations. The restoration of this symmetry is shown to be independent of the geometry of the problem. These results provide a simple and powerful tool for the study of light-matter interactions within the framework of symmetries and conservation laws. We apply such framework to the experimental investigation of helicity transformations in cylindrical nanoapertures, and we find that the transformation is significantly enhanced by the coupling to surface modes, where electromagnetic duality is strongly broken.
\end{abstract}

\maketitle

\section{Introduction}
%%157
Symmetries, both continuous and discrete, are a powerful tool for studying Nature. According to Noether's celebrated theorem \cite{Noether1918}, any continuous symmetry of a non-dissipative system gives rise to a conserved quantity in the dynamic equations. In modern algebraic terms we say that when a system is invariant under the continuous transformation generated by a given operator, the observable represented by that operator is a conserved quantity. For example, rotational and translational invariance are associated with conservation of angular momentum and linear momentum because, as transformations, rotations are generated by the components of angular momentum and translations are generated by the components of linear momentum.

After Einstein founded his theory of space and time on relativistic invariance, the study of symmetries has never left center stage in modern physics. Wigner found that fundamental particles are defined as objects that are invariant under certain transformations: The same principle is currently used for both standard model and beyond standard model physics. Many of the symmetry transformations used in physics have a geometrical nature: translations, rotations, Lorentz boosts. Some others have a more abstract character, separated from geometry, like the isospin symmetry which groups together protons and neutrons into nucleons or the color symmetry of hadrons, which governs the strong force interactions. In light matter interactions, symmetry considerations allow for instance to establish rules for exciting atomic electrons with light. Symmetry reasons are also behind the few exact solutions of Maxwell's equations in inhomogeneous media, like planar multilayers and spheres. Other geometries, like finite cylinders, are also best studied by exploiting their symmetries, as will be shown in the scattering experiments off cylindrical nanoapertures that we present. 

In this paper we will study a non-geometrical symmetry in electromagnetism in detail: electromagnetic duality. Electromagnetic duality is a transformation where the roles of electric and magnetic fields are mixed. Mathematically, the generalized duality transformation of electromagnetic fields is expressed as: 

\begin{equation}
\label{eq:gendual}
\begin{split}
\mathbf{E}&\rightarrow \mathbf{E}_\theta=\cos(\theta) \mathbf{E} - \sin(\theta)\mathbf{H}, \\
\mathbf{H}&\rightarrow \mathbf{H}_\theta=\sin(\theta) \mathbf{E} + \cos(\theta)\mathbf{H}.
\end{split}
\end{equation}
The typical duality transformation, $\mathbf{E}\rightarrow\mathbf{H}$ and $\mathbf{H}\rightarrow-\mathbf{E}$, is recovered by setting $\theta = - \frac{\pi}{2}$. In the absence of charges and currents, (\ref{eq:gendual}) is a symmetry of Maxwell's equations: If the electromagnetic field  $(\mathbf{E},\mathbf{H})$ is a solution of the free space Maxwell equations, then the field $(\mathbf{E}_\theta,\mathbf{H}_\theta)$ is also a solution. In 1965, Calkin \cite{Calkin1965} showed that helicity was the conserved quantity related to such symmetry. 

Helicity is defined \cite[chap. 8.4.1]{Tung1985} as the projection of the total angular momentum $\mathbf{J}$ onto the linear momentum direction $\mathbf{P}/|\mathbf{P}|$, i.e. $\Lambda=\mathbf{J}\cdot\mathbf{P}/|\mathbf{P}|$. In the case of photons \cite[chap. 2.5]{Weinberg1995}, helicity takes the values $\pm1$. It is possible to intuitively understand the meaning of helicity when considering the wave function of the particle in the momentum representation, that is, as a superposition of plane waves. In this representation, helicity is related to the handedness of the polarization of each and every plane wave. Only when all the plane waves have the same handedness is the helicity of the particle well defined. Note that polarization in momentum space and polarization in real space are not the same concept. What Calkin showed is that, as an operator, helicity generates generalized duality transformations in the same way that linear momentum generates translations and angular momentum generates rotations. Since that seminal work, the role of helicity as the generator of generalized duality symmetry transformations for the free space Maxwell's equations has been reported several times \cite{Deser1976,Salom2006,Cameron2012}. 

In 1968, Zwanzinger \cite{Zwanziger1968} extended this free space invariance and conservation law to a material quantum field theory with both electric and magnetic charges. In material systems, the symmetry studied by Zwanzinger seems to be broken by the lack of experimental proof of the existence of magnetic charges. The experimental efforts to find magnetic monopoles are a very active field of research \cite{Milton2006,Bramwell2009,Ladak2010} because, as Dirac demonstrated \cite{Dirac1948}, the mere existence of a single magnetic monopole in the universe would explain the quantization of electromagnetic charge. In Zwanziger's work, transformation (\ref{eq:gendual}) is complemented with a corresponding mixing of electric and magnetic charges and currents. In the absence of magnetic charges, the microscopic Maxwell's equations are no longer invariant under the generalized duality transformations of (\ref{eq:gendual}). This is the current status of duality symmetry in material systems. 

In this article we show that the generalized electromagnetic duality symmetry, broken for the microscopic Maxwell's equations by the absence of magnetic charges, can be restored for the macroscopic Maxwell's equations for material systems characterized by electric permittivities and magnetic permeabilities. The restoration condition for a system composed of different isotropic and homogeneous domains depends only on the materials and is independent of the shapes of the domains. When the system is dual, the helicity of the light interacting with it is preserved.

The geometry independent extension to material systems presented in this article turns the relationship between helicity and duality into a simple and powerful tool for the practical study of light-matter interactions using symmetries and conserved quantities. The practical applicability of our ideas is enabled by a fact of crucial importance: Measurement and preparation of light beams with well defined helicity can be done with very simple optical elements. Armed with this tool, we experimentally investigate helicity transformations in focused light fields that interact with cylindrical nanoapertures in a gold film over a glass substrate. The study of the symmetries of the system allows us to identify the exact reason for the relatively large helicity conversion that we find in the nanoapertures: The coupling of light to surface modes, which, being equal weight superpositions of modes of opposite helicity, strongly break duality symmetry. This result shows the ability of the framework to make both qualitative and quantitative predictions. 

\section{Helicity as the generator of generalized duality transformations in free space}
In our derivations, we will use a harmonic decomposition of the fields and assume a $\exp(-i\omega t)$ dependency with the angular frequency $\omega$. Additionally, we will work in the representation of space dependent vectorial fields, also known as the real representation. This setting is different from those in \cite{Calkin1965} and \cite{Zwanziger1968}, and, although the final result is not new, the derivation in this section sets the stage for the study of the piecewise homogeneous and isotropic case.

The expression of the helicity operator for monochromatic fields in the real representation can be obtained directly from the definition of helicity: 
\begin{equation}
\label{eq:helchiron}
	\Lambda=\frac{\mathbf{J}\cdot\mathbf{P}}{|\mathbf{P}|}=\frac{\left(\mathbf{S}+\mathbf{L}\right)\cdot\mathbf{P}}{|\mathbf{P}|}=\frac{\mathbf{S}\cdot\mathbf{P}}{|\mathbf{P}|}=\frac{\nabla\times}{k}.
\end{equation}
where $\mathbf{S}$ and $\mathbf{L}$ are, respectively, the spin and orbital angular momentum operators, the third equality follows from the orthogonality of $\mathbf{L}=\mathbf{r}\times \mathbf{P}$ and $\mathbf{P}$, and the last one is valid in the real representation because $\mathbf{S}\cdot\mathbf{P}=\nabla\times$ \cite[expr. XIII.93]{Messiah1958} and $|\mathbf{P}|$ is equal to the wavenumber $k$ for monochromatic fields.

Related to the different settings mentioned above, a clarification regarding different definitions of helicity is in order before we start. In \cite{Calkin1965}, helicity appears as an operator in the Fock space representation and in \cite{Zwanziger1968} as an integral involving the electric and magnetic fields and potentials operators. Both of these definitions have since then appeared in the literature several times. Here are the two expressions, in a slightly different notation from that of the original references:
\begin{equation}
\label{eq:helcalkin}
\Lambda = \hbar \int d\kvec \left(a^\dagger_{\kvec +}a_{\kvec +}-a^\dagger_{\kvec -}a_{\kvec -}\right),
\end{equation}
where the Fock space operators $\left(a^\dagger_{\kvec\pm},a_{\kvec\pm}\right)$ create and annihilate photons of definite momentum $\kvec$ and helicity $\pm$. And,

\begin{equation}
\label{eq:helz}
	\Lambda = \frac{1}{2}\int d\mathbf{r}\left(\mathbf{\hat{A}}(\mathbf{r})\cdot\mathbf{\hat{H}}(\mathbf{r})-\mathbf{\hat{C}}(\mathbf{r})\cdot\mathbf{\hat{E}}(\mathbf{r})\right),
\end{equation}
where $\left(\mathbf{\hat{E}},\mathbf{\hat{H}}\right)$ are the electric and magnetic field operators and $\left(\mathbf{\hat{C}},\mathbf{\hat{A}}\right)$ the electric and magnetic potential operators \cite{Zwanziger1968}.

All of the expressions (\ref{eq:helchiron}), (\ref{eq:helcalkin}) and  (\ref{eq:helz}) generate the same fundamental symmetry transformation, albeit in different representation spaces.

We start the derivation by setting convenient units of $\epsilon_0=\mu_0=1$ for the vacuum electric and magnetic constants (thus $c=1$ and $k=\omega$). We can then use (\ref{eq:helchiron}) to write the free space Maxwell equations as:

\begin{equation}
	\label{eq:curl}
\begin{split}
\nabla \times \mathbf{E}=\textrm{i}\,\omega\mathbf{H} &\Rightarrow \mathbf{H}=-\textrm{i}\,\Lambda\mathbf{E} \\
\nabla \times \mathbf{H}=-\textrm{i}\,\omega\mathbf{E} &\Rightarrow \mathbf{E}=\textrm{i}\,\Lambda\mathbf{H}. 
\end{split}
\end{equation}
%93
Equations (\ref{eq:curl}) already reveal that $\Lambda$ is an operator that transforms electric fields into magnetic fields and vice versa. Note that invariance under generalized duality transformation can be interpreted as equivalence between electric and magnetic responses. In the same way that angular momentum generates rotation matrices \cite{Rose1957}, let us use $\Lambda$ as the generator of a continuous transformation parametrized by the real number $\theta$: $D(\theta)=\exp(i\theta \Lambda)$. To obtain an explicit expression for the transformation that $D(\theta)$ performs on the fields, we start by showing that $\Lambda^2$ is the identity operator for Maxwell fields.
\begin{equation}
\label{eq:lambdasqsq}
\Lambda^2 \mathbf{E} = \Lambda \left(\frac{\mathbf{H}}{-i}\right) = \mathbf{E},\
\Lambda^2 \mathbf{H} = \Lambda \left(\frac{\mathbf{E}}{i}\right) = \mathbf{H},
\end{equation}
where the equalities in each equation follow from (\ref{eq:curl}). Since (\ref{eq:lambdasqsq}) is valid for all $\bf{E}$ and $\bf{H}$, we conclude that $\Lambda^2 =I$ for Maxwell fields. Using that $\Lambda^2 =I$, and the Taylor expansion of the exponential, the continuous transformation generated by helicity can be written:
\begin{equation}
\label{eq:dcossin}
D(\theta)=\exp(i\theta \Lambda)=\cos(\theta)I+i\sin(\theta)\Lambda.
\end{equation}
The application of $D(\theta)$ to both electric and magnetic fields reads
\begin{equation}
\begin{split}
\mathbf{E}_\theta&= \left(\cos(\theta) I + i\sin(\theta)\Lambda\right)\mathbf{E},\\
\mathbf{H}_\theta&= \left(\cos(\theta) I + i\sin(\theta)\Lambda\right)\mathbf{H},
\end{split}
\end{equation}
which, after using  (\ref{eq:curl}) again, becomes the well know \cite[chap. 6.11]{Jackson1998} generalized duality transformation of electromagnetic fields written in (\ref{eq:gendual}).

\section{Generalized duality symmetry in piecewise homogeneous and isotropic media}\label{sec:goodstuff}
We will now show that generalized duality symmetry can be restored in the macroscopic Maxwell's equations independently of the shapes of the material domains involved. The macroscopic Maxwell equations are valid whenever the electric and magnetic fields are averaged over many of the atoms or molecules composing the materials. In this way, for most situations, the electromagnetic properties of the materials are determined only by the electric permittivities $\epsilon$ and magnetic permeabilities $\mu$.

We consider an inhomogeneous medium $\Omega$ composed of several material domains with arbitrary geometry. We assume that each domain $i$ is homogeneous and isotropic, and fully characterized by its electric $\epsilon_i$ and magnetic $\mu_i$ constants (we again use $\epsilon_0=\mu_0=1$). In each domain, the constitutive relations are $\mathbf{B}=\mu_i\mathbf{H},\ \mathbf{D}=\epsilon_i\mathbf{E}$, and the curl equations for monochromatic fields read
\begin{equation}
\nabla \times \mathbf{H} = -i\omega \mathbf{D}=-i\omega\epsilon_i \mathbf{E},\
\nabla \times \mathbf{E} = i\omega \mathbf{B}= i\omega\mu_i\mathbf{H}.
\end{equation}
Using $\Lambda=(1/k)\nabla \times$ from (\ref{eq:helchiron}), and $\omega=k_0=k/\sqrt{\epsilon_i\mu_i}$ we obtain
\begin{equation}
\Lambda \mathbf{H} = -i\sqrt{\frac{\epsilon_i}{\mu_i}} \mathbf{E},\ \Lambda \mathbf{E} = i\sqrt{\frac{\mu_i}{\epsilon_i}} \mathbf{H}.
\end{equation}
Note that to arrive at this result, the fact that the wavenumber in each medium is $k=k_0\sqrt{\epsilon_i\mu_i}$ has to be used in the expression of the helicity operator. With this change, we are able to obtain the formal expression of the helicity operator for a material medium, which we could not find in the literature. Now, we can normalize the electric field $\mathbf{E}\rightarrow \sqrt{\frac{\epsilon_i}{\mu_i}}\mathbf{E}$, to show that we can recover the exact form of Maxwell's equations in free space (\ref{eq:curl}). Clearly, the normalization can only be done when all the different materials have the same ratio $\frac{\epsilon_i}{\mu_i}=\alpha\ \forall \ i$. When the normalization is possible, the electromagnetic field equations on the whole medium $\Omega$ are invariant under the generalized duality transformations of (\ref{eq:gendual}).

The remaining question is what happens at the interfaces between the different domains, where the material constants are discontinuous. We now examine the boundary conditions in $\Omega$. At the interfaces between media, the electromagnetic boundary conditions impose the following restrictions on the fields:
\begin{equation}
\label{eq:boundaryconditions}
\begin{split}
	\nhat \times (\mathbf{E_1}-\mathbf{E_2})=0,&\ \nhat \times (\mathbf{H_1}-\mathbf{H_2})=\mathbf{K},\\
\nhat \cdot (\mathbf{D_1}-\mathbf{D_2})={\sigma},&\ \nhat \cdot (\mathbf{B_1}-\mathbf{B_2})=0,
\end{split}
\end{equation}
where $\mathbf{K}$ is the surface current density, $\sigma$ the charge density and $\nhat$ the unit vector perpendicular to the interface. The boundary conditions can be seen as applying point to point to a differential surface area at the interface between the two media \cite[chap. 2.8]{Novotny2006}. Let us choose a particular point $\rvec$ on the interface. Assuming no free charges, i.e $\mathbf{K}=0$ and $\sigma=0$, equations (\ref{eq:boundaryconditions}) may be interpreted as a linear transformation applied to the fields at one medium which results in the fields at the other medium. Using (\ref{eq:boundaryconditions}) and the constitutive relations, the transformation equation reads:
\begin{equation}
\label{eq:T}
\begin{bmatrix} \mathbf{E_2}(\rvec)\\\mathbf{H_2}(\rvec)\end{bmatrix}=
\begin{bmatrix} 
1&0&0&0&0&0\\
0&1&0&0&0&0\\
0&0&\frac{\epsilon_1}{\epsilon_2}&0&0&0\\
0&0&0&1&0&0\\
0&0&0&0&1&0\\
0&0&0&0&0&\frac{\mu_1}{\mu_2}\\
\end{bmatrix}
\begin{bmatrix} \mathbf{E_1}(\rvec)\\\mathbf{H_1}(\rvec)\end{bmatrix},
\end{equation}
where we have oriented our reference axis so that $\nhat=\zhat$. 

On the other hand, the generalized duality transformation (\ref{eq:gendual}) may also be written in matrix form:
\begin{equation}
\begin{bmatrix} \mathbf{E}_\theta\\\mathbf{H}_\theta\end{bmatrix}=\begin{bmatrix} I\cos(\theta)&-I\sin(\theta)\\I\sin(\theta)&I\cos(\theta)\end{bmatrix}\begin{bmatrix} \mathbf{E}\\\mathbf{H}\end{bmatrix}=U(\theta)\begin{bmatrix} \mathbf{E}\\\mathbf{H}\end{bmatrix},
\end{equation}
where $I$ is the $3\times3$ identity matrix. It is a trivial exercise to check that the transformation matrix of (\ref{eq:T}) commutes with $U(\theta)$ if and only if $\epsilon_1/\mu_1=\epsilon_2/\mu_2$. In such case, the fields in each of the two media can be transformed as in (\ref{eq:gendual}) while still meeting the boundary conditions at point $\rvec$. We can now vary $\rvec$ to cover all the points of the interface and repeat the same argument: The fact that $U(\theta)$ does not depend on the spatial coordinates allows to reorient the reference axis as needed to follow the shape of the interface between two media. The derivation is hence independent of the shape of the interface, and we may say that the boundary conditions are invariant under generalized duality transformations when $\epsilon_1/\mu_1=\epsilon_2/\mu_2$.
%the rotation needed to ensure $\nhat=\zhat$ will be different at each point. Nevertheless, due to the crucial fact that $\Lambda$ commutes with the three components of $\mathbf{J}$ \cite[chap. 8.4.1]{Tung1985}, $U(\theta)$ commutes with any rotation, and the same conclusions regarding when $M$ and $U(\theta)$ commute can be reached for all $\rvec$.
The above derivations show that both the equations and the boundary conditions in $\Omega$ are invariant under (\ref{eq:gendual}) when $\frac{\epsilon_{i}}{\mu_{i}}=\alpha \ \forall \ \textrm{domain }i.$ As a conclusion, we can state that independently of the shapes of each domain, a piecewise homogeneous and isotropic system has an electromagnetic response that is invariant under duality transformations if and only if all the materials have the same ratio of electric and magnetic constants: 
\begin{equation}
\label{eq:epsmu}
\frac{\epsilon_{i}}{\mu_{i}}=\alpha \ \forall \ \textrm{domain }i.
\end{equation}
In this case, since helicity is the generator of generalized duality transformations, the system preserves the helicity of light interacting with it.

Our results are in agreement with Bialynicki-Birula's wave equation for photons propagating in a linear, time-independent, isotropic and inhomogeneous medium. In \cite[\S 2]{Birula1996}, he shows that the two helicities of the photon are only coupled through the gradient of $\sqrt{\frac{\mu(\mathbf{r})}{\epsilon(\mathbf{r})}}$.

In the same review \cite[\S 11]{Birula1996}, the author discusses the conservation of helicity in arbitrarily curved spacetime. This notable fact is related to the equivalence of the free space Maxwell's equations on an arbitrary spacetime geometry and the macroscopic equations on a flat spacetime occupied by an anisotropic medium. This equivalence is fundamental in transformation optics \cite[\S 4]{Leonhardt2009}, the theoretical basis for metamaterials. Not all anisotropic media represent spacetime geometries, there is a necessary and sufficient condition for it \cite[\S 4]{Leonhardt2009}: The electric permittivity and magnetic permeability tensors must be equal to each other and equal to a certain function of the spacetime metric $\epsilon^{i,j}=\mu^{i,j}=f(g^{i,j})$ \cite[\S 4]{Leonhardt2009}. At this point, the possible relationship with our results is apparent through the condition $\epsilon^{i,j}=\mu^{i,j}$ and the conservation of helicity discovered by Bialynicki-Birula. Extending the proof in this section to interfaces between anisotropic media could lead to new insights.

Relation (\ref{eq:epsmu}) is often referred to as surface impedance matching condition. It has already been explored in scattering from spheres \cite{Kerker1983} (see appendix \ref{sec:kerker}) and in the context of plasmonic metamaterials \cite{Lee2005}. The connection with helicity preservation was not considered in these references, and, to the best of our knowledge, it has not been considered in any of the references where a relationship as the one in (\ref{eq:epsmu}) is used. The above derivation shows that these particular cases are part of a more general, geometry independent symmetry: generalized electromagnetic duality. 

\begin{figure}[tbp]
 \begin{center}
	\includegraphics[scale=0.9]{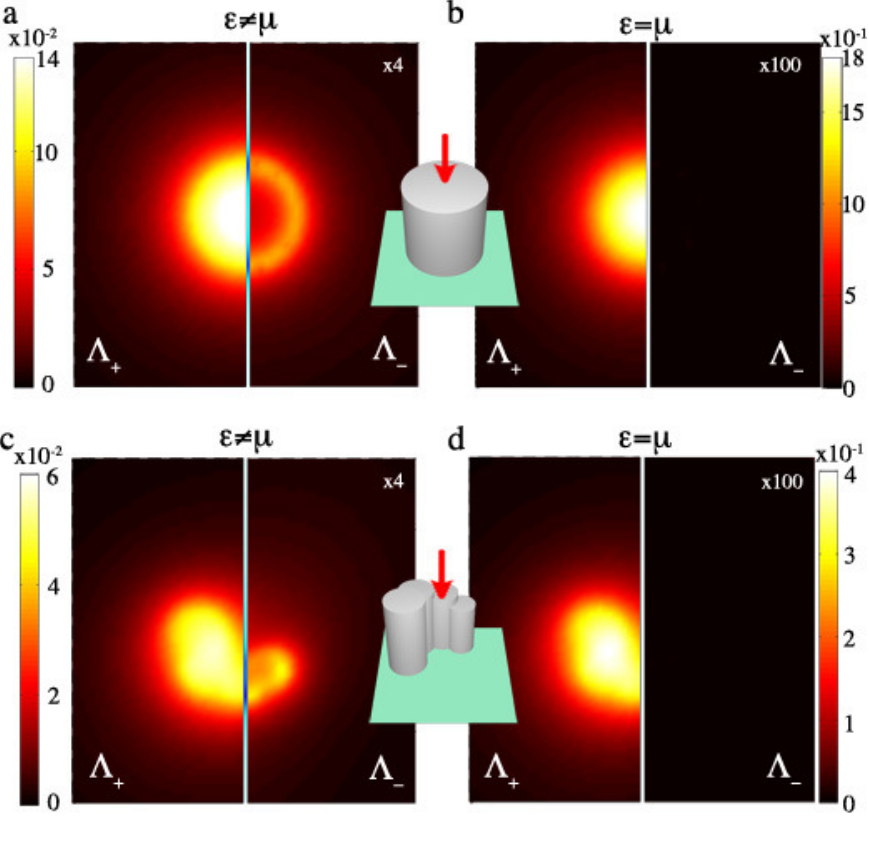}
\end{center}
\caption{(Color) Impact of the different symmetries on the field scattered by two dielectric structures. The upper row shows the intensity calculations for a symmetric cylinder, while the scatterer used for the lower row has a panflute like shape, which lacks all rotational or translational symmetry. In  (a) and (c) the structures are made of silica, while in (b) and (d) we enforced duality symmetry by setting $\epsilon=\mu=\epsilon_{\textrm{glass}}=2.25$. The left half side of each subfigure corresponds to the scattered field with helicity equal to the incident plane wave $\Lambda_+$, while the right half is for the opposite helicity $\Lambda_-$. In order to use the same color scale for all figures, the right half side is multiplied by the factor in the upper right corner. As can be seen, the (lack of) cylindrical symmetry of the structures results in (non-)cylindrically symmetric field patterns, which is consistent with the geometry of each case. On the other hand, both scatterers behave identically with respect to conservation of helicity, which is seen to depend exclusively on the electromagnetic properties of the material. In further simulations we verified that helicity conservation is independent of the angle of incidence. This figure is a clear illustration that cylindrical symmetry is related to conservation of angular momentum, while duality is related to conservation of helicity.}
\label{newfig1}
\end{figure}

In order to illustrate the independence of helicity conservation from geometry we performed numerical simulations. We analyzed the helicity change (Fig. \ref{newfig1}) for two different dielectric structures in free space: A circular cylinder, which is symmetric under rotations along its axis, and a curved panflute like structure without any rotational or translational symmetry. Two versions of each structure were simulated, corresponding to two different materials: the first one would represent silica by setting $\epsilon=\epsilon_{\textrm{glass}}=2.25$ and $\mu=\mu_{\textrm{glass}}=1$. In the second material we enforce duality (\ref{eq:epsmu}), by setting $\epsilon=\mu=\epsilon_{\textrm{glass}}=2.25$. The incident field is a circularly polarized plane wave (i.e. it has well defined helicity) propagating parallel to the axis of the cylinder and the curved surface of the panflute. Fig. \ref{newfig1} shows that helicity is conserved independently of the spatial symmetries, whenever eq. (\ref{eq:epsmu}) is fulfilled, i.e. under conditions of duality symmetry. On the other hand, conservation of angular momentum, resulting in cylindrical symmetry of the scattered fields, is only achieved in the case of the cylinder. 

The most important result of this article is the geometry independent restoration of generalized duality symmetry in the macroscopic Maxwell's equations, turning the relationship between helicity and duality into a practical tool. Helicity and duality can now be added to the toolbox used for the study of light-matter interaction problems within the framework of symmetries and conserved quantities. This tool and framework have already been used to show that the mechanism of optical spin to orbital angular momentum conversion is an inconsistent explanation \cite{FerCor2012b}. The concept of spin to orbit conversion applied to focusing and scattering is masking two completely different physical phenomena related to the breaking of two different fundamental symmetries: transverse translational symmetry in focusing and electromagnetic duality symmetry in scattering. Appendix \ref{sec:kerker} contains another example of the application of our ideas. The unusual scattering effects from magnetic spheres reported in \cite{Kerker1983} are explained in a straightforward manner with the use of duality, helicity and other symmetries and conserved quantities of the system. 
\section{Preparation, measurement and analysis of Maxwell fields with well defined helicity}\label{sm}
Let us now turn to the practical matter of how to prepare and measure electromagnetic helicity in optical experiments. To the best of our knowledge, there has not been any proposal for controlling and analyzing the helicity content of electromagnetic beams. As discussed in the Introduction, the helicity operator provides information about the polarization of the plane wave modes composing the electromagnetic field. From a purely optical perspective, it seems intuitive to use a combination of lenses, waveplates and polarizers to study helicity. This is indeed our approach as can be seen in Fig. \ref{newfig2}(c). The detailed analysis is somewhat involved since it must be done for general (non-paraxial) Maxwell modes. For this reason, we provide the theoretical background needed to understand the structure of a useful set of modes of well defined helicity, and we explain the use of aplanatic lenses and waveplates to control and measure helicity. Additionally, we identify the experimental signature of a scatterer that preserves angular momentum but partially converts helicity. All this background will allow the analysis of our experimental setup and results using symmetries and conserved (or non-conserved) quantities. In our experiments, we illuminate cylindrically symmetric nanoapertures with beams of well defined helicity and analyze the forward scattered field in terms of its helicity content.

\subsection{Electromagnetic modes with well defined angular momentum ($J_z$) and helicity ($\Lambda$): Bessel beams.}\label{sm1}

Typically, modes of well defined helicity, when studied in real space coordinates, contain all the possible polarizations. Let us take a look at a set modes which are very relevant in our experimental setup: Bessel beams. Bessel beams are cylindrically symmetric transverse (i.e. zero divergence) solutions of Maxwell's equations. They have well defined values of energy $H$ and third components of angular and linear momentum $J_z,P_z$. Among the several families of Bessel beams, we will use the one in which they are also eigenstates of helicity, $\Lambda$. Helicity commutes with all the other operators. The set of observables $H,J_z,P_z$ and $\Lambda$ fully determines the electromagnetic field. This kind of Bessel beams have been reported, for example in \cite{Jauregui2005}. A complete derivation can be found in \cite[app. B]{FerCor2012b}.

Below we give the expressions for the two types of vector wave functions $\Chkz$ and $\Dhkz$ having sharp values of $H=k$, $P_z=p_z$, $J_z=m$ and $\Lambda$ equal to $-1$ and $+1$ respectively. We use cylindrical coordinates $[\rho,\theta,z]$ for the spatial variables and the helical basis $[\rhat,\lhat,\zhat]$ for the vectorial character of the fields, where $\lhat=(\xhat + i \yhat)/\sqrt{2}$, $\rhat=(\xhat - i \yhat)/\sqrt{2}$. An implicit harmonic $\exp(-i w t)$ dependence is assumed.

{\small
\begin{equation}
\begin{split}
\label{eq:CD}
&\Chkz(\rho,\theta,z)=
A(z)\exp(i m \theta)\left[ B_+ J_{m+1}(\krho\rho)\exp(i\theta)\rhat+B_- J_{m-1}(\krho\rho)\exp(-i \theta)\lhat+i \sqrt{2}\krho J_m(\krho\rho)\zhat\right],\\
&\Dhkz(\rho,\theta,z)=
A(z)\exp(i m \theta)\left[ B_- J_{m+1}(\krho\rho)\exp(i\theta)\rhat+B_+ J_{m-1}(\krho\rho)\exp(-i \theta)\lhat-i \sqrt{2}\krho J_m(\krho\rho)\zhat\right],\\
\end{split}
\end{equation}
}
where $\krho^2=k^2-\kz^2=p_x^2+p_y^2$, $J_m(\cdot)$ are the Bessel functions of the first kind, the amplitude $A(z)=\sqrt{\frac{\krho}{2\pi}}i^m \exp(i \kz z)\frac{i}{\sqrt{2}}/k$, and $B_{\pm}=(k\pm\kz)$. These modes form a complete orthonormal basis of transverse Maxwell fields.

Note that the structure of the modes with opposite helicity is very similar and it is only distinguished by constants multiplying the different vector components. In the collimated limit, when $\frac{\krho}{k} \rightarrow 0$ ($\kz\approx k$ so that $B_+\rightarrow 2$ and $B_-\rightarrow 0$), both $\Chkz$ and $\Dhkz$ approach pure right circularly (RC) and left circularly (LC) polarized modes respectively. The other polarization components, the opposite circular and the longitudinal $\zhat$ component, are strongly attenuated in the collimated regime. Then, as $\frac{\krho}{k} \rightarrow 0$, the modes in (\ref{eq:CD}) can be approximately written as: 
\begin{align}
\label{eq:Ccollimated}
\Chkz(\rho,\theta,z)&\approx\sqrt{\frac{\krho}{\pi}}i^{m+1}\exp(i(\kz z))J_{m+1}(\krho\rho)\exp(i\theta (m+1))\rhat,\\
\label{eq:Dcollimated}
\Dhkz(\rho,\theta,z)&\approx\sqrt{\frac{\krho}{\pi}}i^{m+1}\exp(i(\kz z))J_{m-1}(\krho\rho)\exp(i\theta (m-1))\lhat.
\end{align}
The error in the intensity that we are making with this approximation is of the order $(p_\rho/k)^2$. This asymptotic property will allow us to prepare and analyze light beams with well defined helicity using simple optical elements. 

\subsection{Preparation and measurement of Maxwell fields with well defined helicity} \label{sm2}

Relations (\ref{eq:Ccollimated}) and (\ref{eq:Dcollimated}) indicate that for collimated light beams, controlling the polarization of the field allows us to control its helicity content. For example, let us consider our experimental setup of Fig. \ref{newfig2}(c). We start with a collimated Gaussian beam with diameter $w=5$ mm. We use a linear polarizer and a quarter waveplate after the laser source in order to obtain a collimated LC polarized Gaussian beam. If we expand such beam with the collimated modes of Eqs. (\ref{eq:Ccollimated}) and (\ref{eq:Dcollimated}), the $\Dhkz$ components would dominate because of the polarization selectivity of the linear polarizer followed by the quarter waveplate. The LC/RC polarization intensity ratio of a Gaussian beam with well defined $\Lambda=+1$ will be of the order $(wk)^{-4}$, which in our case would be of the order of $10^{-19}$. A comparison of this figure with the typical extinction ratios of commercial polarizers,  $10^{-5}$, indicates that the polarizer is the limiting factor when preparing a collimated beam with well defined helicity.

After the polarizer and waveplate, we have prepared an electromagnetic field which is a superposition of only $D$-type Bessel beams. The amplitudes of the $\Dhkz$ for different $m$ and $p_z$ will be given by the shape of our beam. In particular, if the collimated beam is cylindrically symmetric, the value of $m$ controls the azimuthal phase of the field, as seen in (\ref{eq:Dcollimated}). Since our LC Gaussian beam does not contain any azimuthally varying phase, only terms with $m=1$ are possible in its expansion. We have hence prepared a $J_z=+1$, $\Lambda=+1$ light beam. Similarly, in order to project a collimated beam onto states of well defined $\Lambda=\pm1$ it is possible to use  a quarter waveplate and a linear polarizer in two orthogonal settings, as we do in our experiments. 

While preparation and measurement are performed on collimated beams, where the approximations (\ref{eq:Ccollimated}) and (\ref{eq:Dcollimated}) hold, the actual interaction of light with the target may happen in between two microscope objectives. For example, the first one focuses the incident beam onto the target from one side and the second one collects and collimates the output scattered field from the other side. Assuming that the two microscope objectives work as aplanatic lenses \cite{Richards1959}, they will not affect the helicity state of the light beam \cite{Bliokh2011}, and, if perfectly aligned with the incident beam will not affect its angular momentum $J_z$ either: a formal proof of these assertions can be found in \cite[app. C]{FerCor2012b}. Since a lens does not change the energy either, the lenses in our experimental set-up only redistribute the amplitudes of the different $p_z$ modes. This has two important consequences. First, by using simple preparation methods it is possible to illuminate a target with a strongly focused beam possessing well defined $\Lambda$, and second, that whatever helicity change is observed by the measurement apparatus can be attributed to the interaction of the light beam with the target.

\subsection{Conservation of angular momentum and helicity change}\label{sm3}
Let us assume that the target in question is cylindrically symmetric, that is, it preserves $J_z$. Whether it preserves helicity or not depends only on the material properties, as illustrated in Fig. \ref{newfig1}. The transformation of a light beam interacting with such a target can be represented by a transfer matrix between Bessel modes ($\Chkz,\Dhkz$). Since the transformation leaves the angular momentum invariant, the transfer between Bessel modes is only allowed between modes with the same index $m$. If the transformation leaves the helicity invariant, as in the case of the aplanatic lens, $C$-type modes will be transferred to $C$-type modes, and similarly for $D$-type modes. If a cylindrically symmetric target partially converts the helicity of an incident beam, there will be a transfer between $C$ and $D$ modes with the same index $m$. Let us study such case in detail.

Let us say that we start with a $J_z=+1$, $\Lambda=+1$ beam which we focus on a cylindrically symmetric target. If a portion of this beam undergoes an angular momentum conserving helicity transformation, becoming  $J_z=+1$, $\Lambda=-1$, then after collimation, the portion of the beam which has undergone the change can be expanded as a sum of only modes of the type 
\begin{equation}
\label{eq:Ccollimatedm1}
\mathbf{C}_{1\kz}(\rho,\theta,z)\approx\sqrt{\frac{\krho}{\pi}}i^{2}\exp(i(\kz z))J_{2}(\krho\rho)\exp(i2\theta))\rhat,
\end{equation}
while the portion which did not experience a helicity change will necessarily be a sum of only modes of the type:
\begin{equation}
\label{eq:Dcollimatedm1}
\mathbf{D}_{1\kz}(\rho,\theta,z)\approx\sqrt{\frac{\krho}{\pi}}i^{2}\exp(i(\kz z))J_{0}(\krho\rho))\lhat.
\end{equation}

Expressions (\ref{eq:Ccollimatedm1}) and (\ref{eq:Dcollimatedm1}) have been obtained by setting $m=1$ in (\ref{eq:Ccollimated}) and (\ref{eq:Dcollimated}). We see that the angular momentum preserving helicity transfer always leaves an azimuthal phase imprint in the collimated regime. In our case, an original LC polarized beam with no phase singularity transforms into a RC polarized beam with a phase singularity of order two and its corresponding zero of intensity at $\rho=0$ (\ref{eq:Ccollimatedm1}) (Note that $J_l(0)=0$ $\forall \ l\neq 0$). This vortex of charge two is hence a signature of an angular momentum preserving and helicity changing scattering. A projective measurement onto the helicity value opposite to the input one followed by a Charge Couple Device (CCD) camera can be used to look for such a signature.

\section{Helicity transformations in nanoapertures}
In order to investigate helicity changes in light matter interactions, we performed a series of experiments with nanoapertures in a gold film over a glass substrate. The transmission of light through isolated nanoholes was first studied in the seminal paper of Bethe \cite{Bethe1944} and is now understood \cite{Genet2007} to be crucially affected by the effect of localized plasmons and surface modes at the metal-dielectric interfaces. Applications of this kind of nanostructures include optical trapping \cite{Juan2011}, funneling of light \cite{Kwak2004} and reshaping of optical fields \cite{Huang2008}.
In our experiments we observed that helicity is transformed, even though the system we probe is cylindrically symmetric. We thus experimentally verify that helicity is decoupled from angular momentum. We analyze the complete set-up and measurement results only in terms of symmetries and conserved quantities. This methodology allows us to make qualitative and quantitative predictions and, as a result, identify the exact reason for the relatively large helicity conversion in the nanoapertures: The coupling of light to surface modes, which, being linear combinations of modes of opposite helicity, strongly break duality symmetry. 

\begin{figure}[htbp]
 \begin{center}
	\includegraphics[scale=.7]{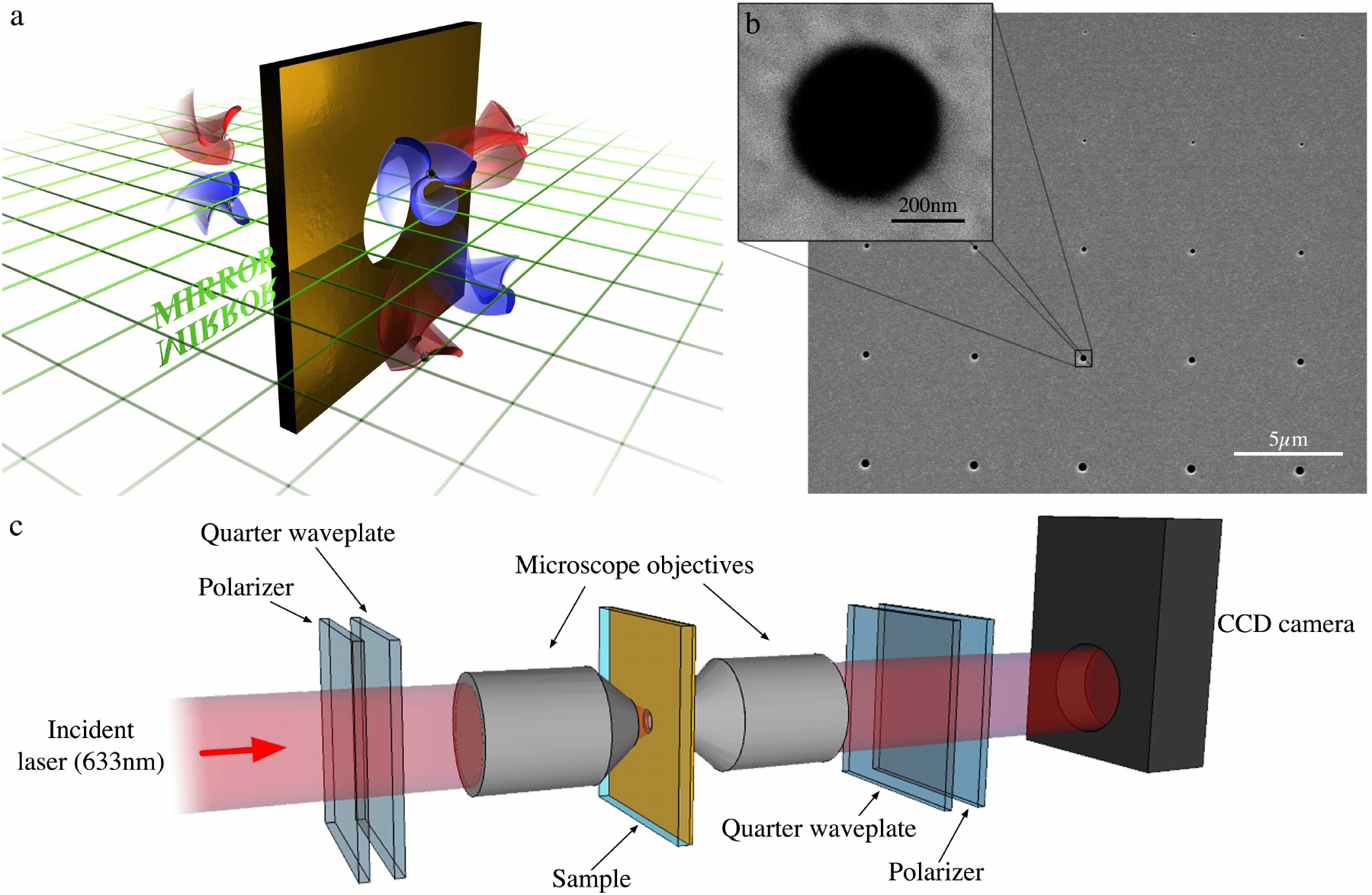}
\end{center}
\caption{(Color) (a), Sketch of the symmetries of the system we are probing. Light with a well defined helicity, represented by a red propeller, impinges on a cylindrically symmetric aperture in a thin metallic film. The output light is analyzed in terms of its helicity content, represented as red propellers when it is the same as the incident helicity, or blue ones with opposite handedness when the helicity is the opposite to the incident one. Note that helicity always flips upon a mirror transformation. This is graphically shown in the sketch by the addition of a mirror plane. (b), Scanning Electron Microscope (SEM) image of the apertures that we tested with our system. We probed a total of 212 different isolated apertures. The separation between apertures was either 5 $\mu$m or 50 $\mu$m and we could not find relevant differences between the two sets. (c), Experimental set-up. An incoming collimated beam is circularly polarized with a set of waveplates and focused to address the isolated nanohole, which is centered with respect to the beam with a nanopositioner. The transmitted light is then collected and analyzed with another set of waveplates and a Charge Coupled Device (CCD) camera.} %138
\label{newfig2}
\end{figure}

Figure \ref{newfig2}(a) sketches the interaction of light and a scatterer consisting of a circular nanoaperture milled in a gold film. Two symmetries of this simple scatterer are noticeable: rotational symmetry around the propagation direction, i.e. the $z$-axis, and mirror symmetry upon reflection through any plane containing the $z$-axis. Rotational symmetry along the $z$-axis implies conservation of the component of angular momentum projected along such axis, i.e. $J_z=\mathbf{\hat{z}}\cdot\mathbf{J}$. All the machinery for the preparation, measurement and analysis of helicity, explicitly developed in section \ref{sm1} for cylindrically symmetric targets, is fully applicable to our experimental setup.

\subsection{Experimental setup, methods and measurements}
%188
The system we have experimentally tested consists of a set of isolated nanoapertures of different sizes, which were milled with a focused ion beam in a gold layer of $200$ nm, deposited on top of a $1$ mm thick glass substrate. The diameters of the nanoapertures ranged from $150$ to $580$ nm.  Using a Scanning Electron Microscope (SEM), we carefully measured the diameters of all the apertures and checked that all the apertures had aspect ratios between $0.95$ and $1$. We probed the nanoapertures with a continuous wave laser with a wavelength of $\lambda_0=633$ nm. The preparation of the probing beam was done as follows. First, we collimated the laser beam and then we used a set of linear polarizers and waveplates to ensure a left circularly polarized light beam. As explained in section \ref{sm2}, when this collimated field is decomposed in modes of well defined $J_z$ and $\Lambda$, the components with $J_z=1$, $\Lambda=1$ are overwhelmingly dominant. This collimated helicity field was subsequently focused with a microscope objective with a numerical aperture of NA=0.5. Since the transformation of an aplanatic lens preserves helicity, we were able to generate a focused electromagnetic field with a well defined helicity. The focused field was then allowed to interact with one of the isolated nanoapertures. We carefully positioned the nanoaperture on the symmetry axis of our optical system by means of a set of piezo-stages.  Subsequently, the scattered light was collected and collimated with another microscope objective of NA=0.9. Once again, this lens did not affect the helicity of the beam, and as such, after collimation, we were able to analyze the helicity with another set of waveplates and polarizers, obtaining two very different spatial profiles for fields with $J_z=1$, $\Lambda=1$ and $J_z=1$, $\Lambda=-1$. The light was detected with a CCD camera.

\begin{figure}[tbp]
 \begin{center}
	\includegraphics[scale=1]{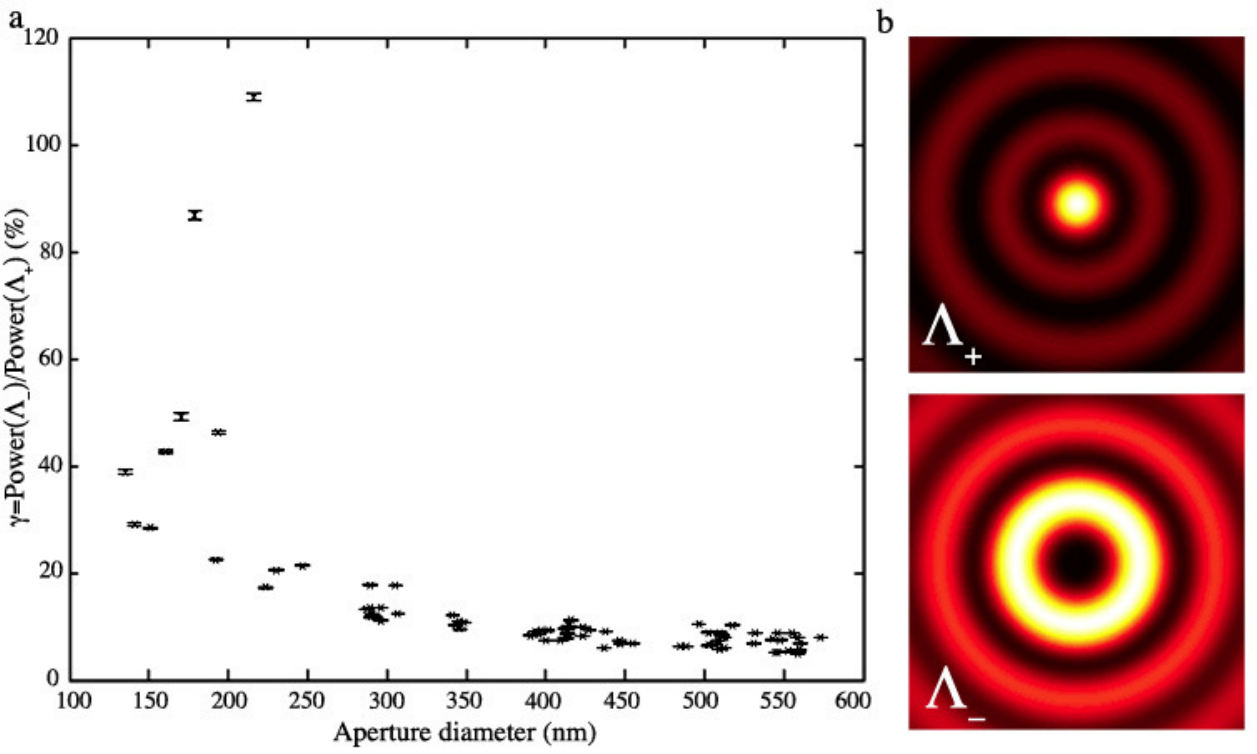}
\end{center}
\caption{(Color) Helicity transformations through nanoapertures. (a), Dependence of the ratio of transmitted helicities on the aperture size. All points correspond to highly symmetrical apertures, the sizes of which were measured with an SEM.  (b), Numerically calculated spatial pattern for the direct and transformed helicity fields transmitted through a cylindrical aperture of $300$ nm} %46
\label{newfig3}
\end{figure}

Our results show that there is always a helicity transformation in the transmitted light. This can be seen in Fig.~\ref{newfig3}(a), where we plot the power ratio between the two transmitted helicities, $\gamma$, as a function of the aperture sizes. The smallest conversion we measured was for the largest holes, $\gamma_{580}=0.08\pm0.02$. In contrast, the helicity transformation measured in the same sample through the glass alone, corresponding to an infinite aperture, is $\gamma_{\inf}\approx10^{-3}$. This conversion value can be due to left over gold nanoparticles distributed over the glass surface. Also, the helicity transformation by the focusing and collimating lenses alone was even smaller, of the order of $\gamma_{\textrm{lens}}\approx10^{-4}$, which is consistent with the fact that perfect aplanatic lenses should preserve helicity.  In Fig.~\ref{newfig3}(b), we display the typical spatial patterns for the two output helicities of the light scattered from a perfect cylindrical aperture, as calculated with a semi-analytical method \cite{FerCor2011}.  We numerically checked that this output conserves angular momentum but, as can be seen, breaks helicity conservation (duality symmetry). The spatial shape and optical vortex content of beams that have undergone an angular momentum preserving but helicity flipping interaction is explained in section \ref{sm3}.

\begin{figure}[tbp]
 \begin{center}
	\includegraphics[scale=.95]{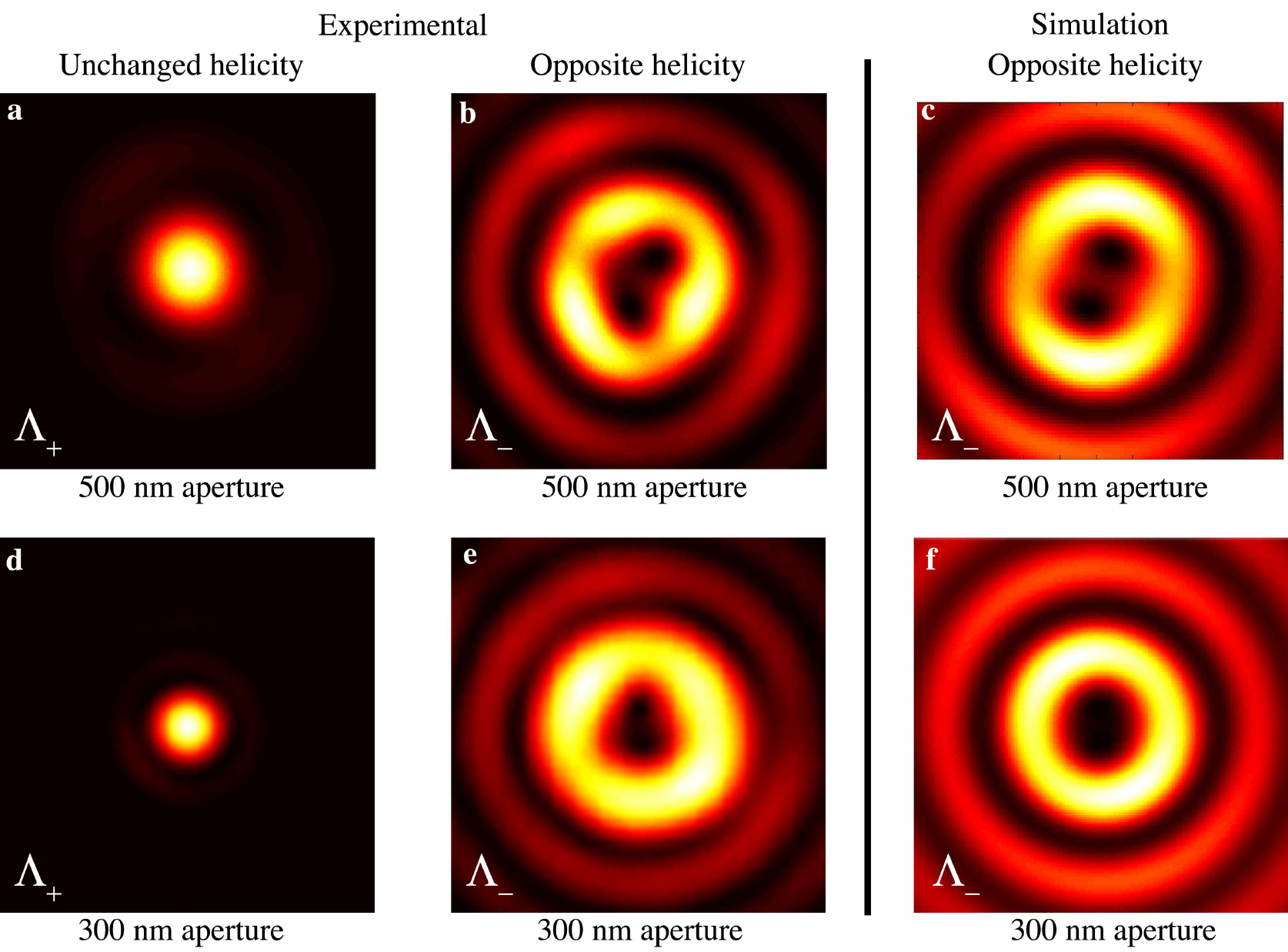}
\end{center}
\caption{(Color) Projective measurement of helicity. Upper (lower) row shows the results for an aperture of $500$ nm ($300$ nm). (a) and (d), Experimental results showing the transmitted light with helicity identical to the incident light. (b) and (e), Experimental results of transmitted light with opposite helicity. The cylindrical asymmetries are due to the finite rejection power of the polarizer. This is clearly seen with the numerically simulated patterns, (e) and (f), where the asymmetry appears only after including the experimental parameters of the used waveplates.}%84 
\label{newfig4}
\end{figure}

%113
In order to test that our experimental results are consistent with helicity changes with conservation of angular momentum, we analyze the CCD images. In Fig.~\ref{newfig4} we show typical experimental results and their comparison with numerical calculations for two different aperture sizes. In the left column (Figs.~\ref{newfig4}(a) and (d)) we show the components of the output field with the same helicity as the input, $\Lambda_+$. The observed field pattern is a typical Airy pattern arising from the subwavelength dimensions of the nanoaperture and the finite numerical aperture of the collection microscope objective, as expected from Fig.~\ref{newfig3}(b). On the other hand, the central column (Fig.~\ref{newfig4}(b) and (e)) shows the field with opposite helicity, $\Lambda_-$. The absence of singularities in Figs.~\ref{newfig4}(a) and (d), and the presence of two singularities in Fig.~\ref{newfig4}(b) and (e), are fully consistent with the results of section \ref{sm3}.

%340
The differences between our experimental results and the ideal case of Fig.~\ref{newfig3}(b) for the helicity transformed transmission are due to the finite extinction ratios of our polarizers. We now discuss this assertion. First of all, electromagnetic modes with well defined $J_z$ have to be cylindrically symmetric. Additionally, as we show in section \ref{sm2}, collimated beams with well defined angular momentum and helicity always present a phase singularity of order $q=J_z-\Lambda$ in the dominant polarization, i.e. the wavefront of the field is twisted around the center $q$ times $2\pi$. In our case, this translates to a smooth phase front, with no phase singularity, for the directly transmitted helicity ($q=1-1=0$) and a second order singularity for the helicity transformed mode ($q=1-(-1)=2$); which is consistent with the experimental images of Fig \ref{newfig4}. In practice, we could not avoid a small leakage from the direct helicity to the transformed helicity. As a result of such a superposition, the second order singularity splits into two singularities of order $1$. Thus, the intensity pattern is no longer cylindrically symmetric. In order to prove this point, we show in Figs.~\ref{newfig4}(c) and (f) the coherent superposition of numerically calculated images of unmixed modes (as those of Fig.~\ref{newfig3}(b)), with relative amplitudes given by the extinction ratios of our set of polarizers, $5\times 10^{-5}$. Given that the leakage through the waveplates is the same for all apertures, one would expect that its effect should be smaller for those scatterers with a higher $\gamma$. This is indeed the case seen in Fig.~\ref{newfig4}, which shows a larger spatial separation of the singularities for the larger aperture. The helicity transfer for the $500$ nm aperture, $\gamma_{500}=0.07\pm{0.01}$, is a factor $2.3$ smaller than for the $300$ nm aperture, $\gamma_{300}=0.16\pm{0.03}$. The only remaining free parameter is the relative phase between the two modes, the effect of which is to rotate the whole pattern around its center. By comparing the experimental results with the calculated pattern we can infer that the relative phase between the modes was ${\pi}/{3}$, for the $500$ nm aperture (Fig.~\ref{newfig4}(c)), and $0$ for the $300$ nm (Fig.~\ref{newfig4}(f)). We conclude that our measurements are consistent with the fact that in our system, the angular momentum is conserved, but helicity is not.

\subsection{Analysis of the experimental measurements}

According to the ideas presented in this paper, the observed helicity change implies that electromagnetic duality is broken in our samples. In order to identify the exact mechanism of duality breaking, it is important to first consider the multilayer system air-glass-gold-air without the nanoaperture. Considering condition (\ref{eq:epsmu}), duality is obviously broken by just the multilayer alone, but the helicity transfer in the absence of the nanoaperture has been numerically shown to be around $10^{-6}$ for our collection objective. This value is one order of magnitude smaller than the typical measurement noise, which is related to the polarizer extinction value of $5\times10^{-5}$. The experimental observation of much higher transformation ratios must hence be tied to the nanoapertures. 

The transmission and reflection in a multilayer system is best studied using plane waves. For a single plane wave with momentum $\mathbf{{k}}$, the two helicity states are the two states of circular polarization, and can be obtained by linear combination of its $\mathbf{s}$ (transverse electric) and $\mathbf{p}$ (transverse magnetic) components: $\mathbf{s}\pm\mathbf{p}$. See \cite[app. A]{FerCor2012b} for a general derivation of this relationship, which also applies to multipolar fields and Bessel beams. Different scattering coefficients for $\mathbf{s}$ and $\mathbf{p}$ will hence mix the two helicity modes. This idea has been recently applied to the analysis of resonances in spheres \cite{Zambrana2012}. If the particular multilayer presents any resonance, for either $\mathbf{s}$ or $\mathbf{p}$, the helicity transfer will be enhanced in its vicinity: A pure $\mathbf{s}$ or pure $\mathbf{p}$ mode is an equal weight combination of the two helicities, and hence strongly breaks helicity conservation (duality symmetry). Our system indeed presents two resonances for non-propagating modes. These two resonances are related to surface modes on the metallic surfaces. It is well known that through the scattering of the nanoaperture, the incident field can excite surface plasmon polaritions (SPP) at the interface. Since SPPs are $\bf p$-polarized waves \cite{Maier2007}, this produces an asymmetric response of the SPP with regard to the transmitted $\bf s$ and $\bf p$-polarized components. This additional SPP induced electromagnetic asymmetry dramatically enhances the helicity transfer, even in the propagating modes, and makes it experimentally detectable. According to this explanation, the helicity transfer should increase for modes in the proximity of the plasmon resonance, i.e. for large transversal momenta. This actually explains the trend in Fig. \ref{newfig2} that smaller holes present a larger $\gamma$ value: smaller holes have a higher coupling to large transversal momenta, and in particular to the surface modes. We can then conclude that the most important role of the nanoaperture in the helicity transfer is to couple the incident field to the resonances and, in particular, to the surface plasmon polaritons.

Very similar experimental setups and measurements have been previously reported. Starting with the work of Gorodetski {\em et al}. \cite{Gorodetski2009}, we can also find \cite{Vuong2010} and very recently \cite{Chimento2012}. In all these cases, the results have been analyzed by means of the spin ($\mathbf{S}$) to orbital ($\mathbf{L}$) angular momentum conversion mechanism. As already mentioned, \cite{FerCor2012b} shows the inconsistency of the spin to orbital explanation, whose root is the fact that only the components of $\mathbf{J}=\mathbf{S}+\mathbf{L}$ are the generators of rotations for the electromagnetic field. The components of $\mathbf{L}$ and $\mathbf{S}$, when separately considered, do not generate meaningful symmetry transformations of the electromagnetic field. This is reflected by the fact that the application of the operator corresponding to any of their components, for instance $L_z$ or $S_z$, breaks the transversality of the fields \cite{VanEnk1994}, \cite[page 50]{Cohen1997}. In this paper, we have used the generator of rotations along the $z$ axis $J_z$, and the generator of generalized duality transformations $\Lambda$, to analyze our experimental results from the point of view of symmetries and conserved quantities. This framework has allowed us to identify the root cause of the relatively high helicity change by means of qualitative and quantitative considerations. Spin to orbit angular momentum conversion, due to its inconsistent nature, does not allow this level of understanding. The analysis contained in this section explains the experimental observations of the above given references, including some of the cases in \cite{Gorodetski2009} and \cite{Vuong2010} where the apertures were non-cylindrical.

\section{Conclusion}
 In this article, we have shown that the restoration of generalized duality symmetry is possible for the macroscopic Maxwell's equations, even though the microscopic equations are rendered asymmetric by the empirical absence of magnetic charges. The restoration of the symmetry is independent of the geometry of the problem: a system made of piecewise isotropic and homogeneous domains of different materials characterized by electric and magnetic constants $(\epsilon_i,\mu_i)$ is invariant under generalized duality transformations if and only if $\frac{\epsilon_i}{\mu_i}=\alpha \ \forall \ \textrm{media } i$. This result is independent of the shapes of the domains. With this result, the known relationship between helicity and generalized duality transformations, namely that the former is the generator of the latter, is turned into a simple and powerful tool for the practical study of light-matter interactions using symmetries and conserved quantities. Armed with it, we have experimentally investigated helicity transformations in focused light fields that interact with cylindrical nanoapertures in a gold film over a glass substrate. Analyzing the results by means of symmetries and conserved quantities, including duality and helicity, we have been able to conclude that the role of the nanoapertures is to allow light to couple to resonant modes of the system where duality is strongly broken: this is what renders the helicity transformation effect observable in our experimental set-up. In our experiments, we have used a fact of crucial practical importance: Measurement and preparation of beams with well defined helicity can be done with very simple optical elements.

We are confident that we are proposing a useful tool. In the article, we have shown how to apply it to an experiment and gain valuable insight with it. In appendix \ref{sec:kerker} we use it to explain some unsual scattering effects in a straightforward way. In \cite{FerCor2012b}, it allowed to proof the inconsistency of the concept of optical spin to orbital angular momentum conversion in focusing and scattering, and to propose a substitute framework based on helicity. Additionally, we are preparing a manuscript were the role of duality symmetry in molecular optical activity is identified.

Our results may be useful in other fields. For example, they may prove important in the field of metamaterials and transformation optics \cite{Chen2010}, which is dramatically extending the range of wavelengths where effective electric and magnetic constants can be engineered. The transfer of helicity between light and matter remains an open line of research, which could have importance in the fields of plasmonics and ``spintronics'' \cite{Valenzuela2006}, where the control of the helicity of electrons is crucial. Finally, it can be seen that the same tools we have developed here can be successfully used to explain recently reported effects in electron beams \cite{Karimi2011}. This parallelism is an encouraging sign towards the possibility of simulating particle interactions on an optical table \cite{Gerritsma2010}. 
\appendix
\section{Scattering effects for magnetic spheres}\label{sec:kerker}
Consider the unusual scattering effects for magnetic spheres reported by Kerker \cite{Kerker1983}. One of them refers to the fact that a plane wave impinging on a vacuum embedded sphere with $\frac{\epsilon}{\mu}=1$ does not produce any backscattered field (at a $180$ degrees scattering angle). This effect, which has been referred to as an anomaly \cite{Nieto2011}, can be easily understood using our results. Let us take as incident field a circularly polarized plane wave with is momentum aligned along the $z$ axis. Its angular momentum is also aligned with the $z$ axis and, in natural units of $\hbar=1$, equal to $\pm 1$ depending on the handedness of the polarization. We now know (equation (\ref{eq:epsmu})) that, based only on the properties of the materials, helicity has to be preserved in the interaction between the plane wave and a dual sphere.  Let us now assume that there exist a component of backscattered field at $180$ degrees, that is, a plane wave whose linear momentum is the negative of the linear momentum of the incident plane wave. Recalling that $\Lambda=\mathbf{J}\cdot\mathbf{P}/|\mathbf{P}|$, we see that, to preserve helicity, the angular momentum of the backscattered plane wave must also change sign with respect to the angular momentum of the incident plane wave. But such change is impossible: the rotational symmetry of the sphere implies that angular momentum is preserved in all axes, in particular along the axes shared by the incident and the backscattered plane waves. As a consequence, the backscattering amplitude must be zero. Since this argument applies independently to both circular polarizations, the backscattering gain will be zero for any polarization of the incoming plane wave. What is sometimes referred to as the first Kerker condition is hereby explained. Interestingly, zero backscattering from dual objects has already been the object of investigation by the radar community \cite{Lindell2009}: in these works, the connection between duality and helicity is not recognized.

In the same paper, Kerker finds that upon scattering off a vacuum embedded sphere with $\frac{\epsilon}{\mu}=1$, the state of polarization of light is preserved independently of the scattering angle. The root cause of such interesting phenomenon is the simultaneous invariance of the system with respect to generalized duality transformations, due to the materials, and any mirror operations through planes containing the origin of coordinates, due to the geometry. In the helicity basis, the 2x2 scattering matrix between an incident and a scattered plane wave \cite[chap. 3]{Weinberg1995} must be diagonal because of helicity preservation. Additionally, it must also preserve the linear polarizations parallel and perpendicular to the plane containing the two plane wave momentum vectors, because a mirror operation across such plane leaves the sphere and both momentum vectors invariant. Using then that helicity flips with mirror operations, it can be easily shown that all the 2x2 scattering matrices are indeed diagonal and hence preserve the state of polarization between any pair of incident and scattered plane waves.

%\bibliographystyle{unsrt}
%\bibliography{ivanfcrefs}

{\bf Acknowledgements}
This work was funded by the Australian Research Council Discovery Project DP110103697 and the Centre of Excellence for Engineered Quantum Systems (EQuS). G.M.-T is also funded by the Future Fellowship program (FF)
\end{document}